\documentclass[twocolumn,showpacs,superscriptaddress,amsmath,amssymb,pre]{revtex4}

\usepackage{graphicx}
\usepackage{times}

\newcommand{\aion}{a}
\newcommand{\nion}{n_\mathrm{ions}}
\newcommand{\Nion}{N_\mathrm{ions}}

\newcommand{\kB}{k}

\newcommand{\beq}{\begin{equation}}
\newcommand{\eeq}{\end{equation}}
\newcommand{\bea}{\begin{eqnarray}}
\newcommand{\eea}{\end{eqnarray}}
\newcommand{\req}[1]{Eq.\ (\ref{#1})}

\newcommand{\PR}[2]{Phys.\ Rev. #1 \textbf{#2}}
\newcommand{\PRL}[1]{{Phys.\ Rev.\ Lett.} \textbf{#1}}

\begin{document}

% ----------   FRONT MATTER:  ------------

\title{Addendum to ``Equation of state of classical Coulomb plasma mixtures''}

\author{A. Y. Potekhin}\email{palex@astro.ioffe.ru}
    \affiliation{Ioffe Physical-Technical Institute,
     194021 St.\ Petersburg, Russia}
    \affiliation{Ecole Normale Sup\'erieure de Lyon,
     CRAL (UMR CNRS No.\ 5574),
     69364 Lyon Cedex 07, France}

\author{G. Chabrier}\email{chabrier@ens-lyon.fr}
    \affiliation{Ecole Normale Sup\'erieure de Lyon,
     CRAL (UMR CNRS No.\ 5574),
     69364 Lyon Cedex 07, France}

\author{A. I. Chugunov}
    \affiliation{Ioffe Physical-Technical Institute,
     194021 St.\ Petersburg, Russia}

\author{H. E. DeWitt}
\affiliation{Lawrence Livermore National Laboratory,
 P. O. Box 808, Livermore, CA 94550, USA}

\author{F. J. Rogers}
\affiliation{Lawrence Livermore National Laboratory,
 P. O. Box 808, Livermore, CA 94550, USA}

\date{\today}

\begin{abstract}

Recently developed analytic approximation for the equation of
state of fully ionized nonideal electron-ion plasma mixtures 
[Potekhin et al., Phys.\ Rev. E \textbf{79}, 016411 (2009)], which
covers the transition between the weak and strong Coulomb
coupling regimes and reproduces numerical results obtained in the
hypernetted chain (HNC) approximation, is modified in order to
fit the small deviations from the linear mixing in the strong
coupling regime, revealed by recent Monte Carlo simulations. In
addition, a mixing rule is proposed for the regime of
weak coupling, which generalizes post-Debye density corrections
to the case of mixtures and numerically agrees with the HNC
approximation in that regime.

\end{abstract}


% 05.70.Ce: Thermodynamic functions and EOS
% 52.25.Kn: Thermodynamics of plasmas
% 52.27.Gr: Strongly-coupled plasmas

\maketitle

%%  SECTION   ---------------------------
\section{Introduction}
\label{sect:intro}

The high accuracy of the linear mixing rule (LMR) for
multicomponent strongly coupled Coulomb plasmas has been
confirmed in a number of papers 
\cite{HV76,HTV77,BHJ79,ChabAsh,DWSC96,DWS99,DWS03}. Nevertheless, the
accuracy of modern Monte Carlo (MC) calculations allows one to
reveal certain deviations from the LMR for the Coulomb
energy $U$ of binary
ionic mixtures (BIM). On the other hand, for weakly coupled plasmas  the
Debye-H\"uckel (hereafter DH) formula is applicable instead of the LMR. 
Several terms in the density expansion of $U$ beyond 
the DH approximation were obtained 
by Abe \cite{Abe} and by Cohen \& Murphy \cite{CohenMurphy}
(hereafter ACM) in the one-component plasma (OCP) case.

In Ref.~\cite{BHJ79}, deviations from the LMR for BIM were studied in
the hypernetted chain (HNC) approximation and fitted by Pad\'e
approximants. 
In Ref.~\cite{ChabAsh}, the LMR was confirmed
by HNC method for \emph{polarizable} background of partially
degenerate electrons.
In Ref.~\cite{DWSC96}, deviations from the LMR for strongly coupled BIM
were studied using both HNC and MC techniques. The corrections to 
the LMR for $U$ were found to be of the same
order of magnitude for HNC and MC, but numerically different;
in particular, it does not depend on 
the mean ion Coulomb coupling parameter
$\Gamma$ according to
HNC results,
but decreases as function of $\Gamma$ in MC simulations.
These results were confirmed in Ref.~\cite{DWS03},
where an analytic fit to the calculated corrections was 
suggested.
The fitting formulae of Refs.~\cite{BHJ79,DWS03} are applicable
only at $\Gamma\geq1$; in particular they do not reproduce
the DH limit at $\Gamma\to0$ (besides, the fit 
parameters in \cite{BHJ79} are given only
for 5 fixed ionic charge ratios from 2 to 8).

In Ref.~\cite{PCR09}, HNC calculations of BIM and
three-component ionic mixtures (TIM) were performed in a wide
range of values of 
$\Gamma$,
charge ratios, and partial densities of the ion components, 
and a parametric formula was suggested to
fit the \emph{fractional} differences between the LMR and
calculated plasma energies at any $\Gamma$ in liquid
multicomponent plasmas. It recovers the DH formula at
$\Gamma\ll1$ and gives a vanishing fractional difference from the
LMR at $\Gamma>1$.

However, in the regime of \emph{strong} coupling, the accuracy of
the HNC method (typically a few parts in 1000, for $U$)  is not
sufficient to reproduce the values of the energies of mixtures at
the precision level needed to study deviations from the LMR
(see, e.g., \cite{DWSC96}). Indeed, according to
Refs.~\cite{BHJ79,DWSC96,DWS03}, these deviations are  typically of the
order of a few $\times(10^{-3}$\,--\,$10^{-2})\, \kB T$ per ion (where $\kB$
is the Boltzmann constant),
while $U\sim-\Gamma \kB T$ per ion at $\Gamma\gg1$.

In this brief report, we suggest two improvements
for analytic treatment of ion mixtures.
First, we introduce a mixing rule for
weakly coupled plasmas, which provides an extension
of the ACM formula to the 
case of ion mixtures and agrees with HNC
results up to the values 
of the Coulomb coupling parameter $\Gamma\approx0.1$
(whereas the DH approximation becomes inaccurate
at $\Gamma\gtrsim0.01$).
Second, using MC simulations of strongly coupled
liquid BIM, supplementary to those already published
in \cite{DWSC96,DWS99,DWS03}, we suggest a modified version of the
formula \cite{PCR09}, which maintains the accuracy of the
previous fit at intermediate and weak coupling, but delivers
consistency with the MC data for strongly coupled Coulomb
liquids.

In Sec.~\ref{sect:basic} we introduce basic notations and
formulae; in Sec.~\ref{sect:CM} we propose a
mixing rule applicable at weak coupling; in
Sec.~\ref{sect:fit} we present a fitting formula for the internal
energy of mixtures, applicable
in the entire domain of $\Gamma$ values for weakly
and strongly coupled classical Coulomb gases and liquids; and 
in Sec.~\ref{sect:concl} we summarize the results.

% --------------------------------------------------------
\section{Basic equations}
\label{sect:basic}

Let $n_e$ be the electron number density and
$n_j$ the number density of ion species
with charge numbers $Z_j$ ($j$=1,2,\ldots).
The total number density of ions is $\nion = \sum_j n_j$.
The electric neutrality implies $n_e = \langle Z \rangle \nion$.
Here and hereafter the angular brackets denote
averaging with statistical weights proportional to $n_j$:
\beq
   \langle Z \rangle \equiv \sum_j x_j Z_j ,
\quad\mbox{where~}x_j\equiv \frac{n_j}{\nion}.
\label{mean}
\eeq

The strength of the Coulomb interaction of ion
species $j$ is characterized by the Coulomb coupling parameter,
defined (in CGS units) as
$ \Gamma_j =
        (Z_j e)^2/a_j   \kB T = \Gamma_e Z_j^{5/3},
$
where
$a_j  = a_e Z_j^{1/3}$
is the ion sphere radius, 
$
    \Gamma_e \equiv { e^2 }/{ a_e  \kB T},
$
and $a_e\equiv (4\pi n_e/3)^{-1/3}$.
In other words, partial coupling parameters $\Gamma_j$ and
ion sphere radii $a_j$ are defined to be those of the OCP
of ions of the $j$th kind at the same electron density
$n_e$ as in the considered multicomponent plasma.
The Coulomb coupling in the mixture of different ions
is conventionally characterized by the
average coupling parameter
$
   \Gamma = \Gamma_e \langle Z^{5/3} \rangle.
$

\begin{figure}[t]
\includegraphics[width=.99\columnwidth]{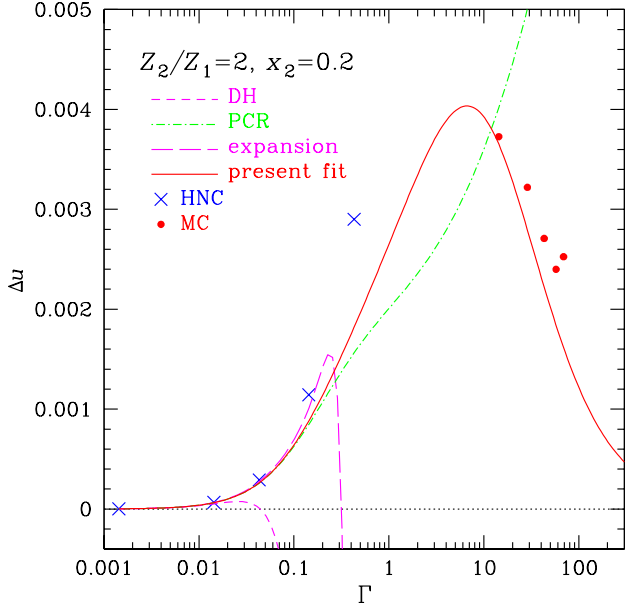}
\caption{\label{fig:umixmc2b}(Color online)
Correction
to the LMR $\Delta u=u-u_\mathrm{LM}$ 
as a function of $\Gamma$
for BIM with $Z_2/Z_1=2$, $x_2=0.2$.
HNC (crosses) and MC (dots) data 
are compared to the DH
approximation (short-dashed lines), the modified ACM
approximation (\ref{NLM}) (long-dashed lines), the fit from \cite{PCR09}
(dot-dashed lines), and the present fit (\ref{fit})
(solid lines).}
\end{figure}

\begin{figure}[t]
\includegraphics[width=.99\columnwidth]{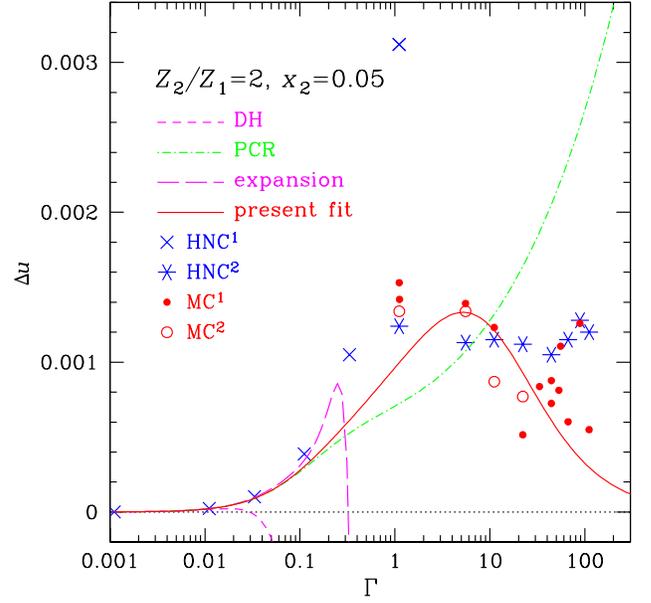}
\caption{\label{fig:umixmc2c}(Color online)
$\Delta u=u-u_\mathrm{LM}$ 
as a function of $\Gamma$ for
BIM with $Z_2/Z_1=2$, $x_2=0.05$. Here crosses (HNC$^1$) correspond to 
$\Delta u$ obtained from the HNC data using
the OCP fit
from \cite{PC00}
for calculation of $u_\mathrm{LM}$,
and
asterisks (HNC$^2$) correspond to $\Delta u$
from Ref.~\cite{DWSC96}, where \emph{both} $u$ and $u_\mathrm{LM}$
are based on the HNC results.
Dots (MC$^1$)
correspond to $\Delta u$
calculated from the recent MC data for $u$,
and $u_\mathrm{LM}$ calculated from the OCP fit \cite{PC00},
while circles (MC$^2$) represent MC data \cite{DWSC96} for
$\Delta u$.}
\end{figure}

A common approximation for the Coulomb contribution to
the internal energy of a strongly coupled ion 
mixture is the LMR:
\beq
u_\mathrm{LM}(\Gamma) 
= \sum_j x_j u(\Gamma_j,x_j=1)\,,
\label{LMR}
\eeq
where $u\equiv U/\Nion \kB T$
is the reduced Coulomb energy, 
$\Nion$ is the total number of all ions,
and the
subscript ``LM'' denotes
the linear-mixing approximation. Obviously,
the LMR has the same form for the Coulomb contribution
to the reduced Coulomb free energy
$f\equiv F/\Nion \kB T$.

When the Coulomb interaction is 
sufficiently weak compared to the thermal
energy, 
then the DH approximation can be applied:
$
   u_\mathrm{DH} = \langle q^2\rangle/\kB Tr_D,
$
where $\langle q^2\rangle$ is the mean squared charge of
the considered mixture and 
$r_D=(\kB T / 4\pi\nion\langle q^2\rangle)^{1/2}$ 
is the Debye radius.
For the model of ions in the ``rigid'' electron background,
applicable if the electrons are extremely strongly degenerate,
$\langle q^2\rangle=e^2\langle Z^2\rangle$,
whereas in the case of completely nondegenerate electrons,
using our definition (\ref{mean})
of averaging over the \emph{ion} species
and taking into account the neutrality condition, we have
$\langle q^2\rangle=e^2(\langle Z^2\rangle+\langle Z\rangle)$.

In this paper we consider the model of rigid electron
background, but extension to the case of
compressible background is possible by 
adjusting the parameter $\delta$ in \req{delta} below,
according to the expression for $\langle q^2\rangle$.
In Ref.~\cite{PCR09} this extension
was shown to be compatible with numerical HNC data \cite{ChabAsh}
for ion mixtures with allowance for electron polarization.

% --------------------------------------------------------
\section{Weakly coupled ion mixtures}
\label{sect:CM}

For a OCP at $\Gamma\ll1$, a cluster expansion yields
\cite{Abe,CohenMurphy}
\bea
   u &=& 
            -\frac{\sqrt{3}}{2}\,\Gamma^{3/2}
            -3\Gamma^3\,\left[\frac38\ln(3\Gamma)
            + \frac{C_E}{2}-\frac13\right]
 \nonumber\\&&
            -\Gamma^{9/2}(1.6875\,\sqrt{3}\,\ln\Gamma - 0.23511)
            +\cdots,
\label{CM}
\eea
where $C_E=0.57721\ldots$ is the Euler constant.
Here, the first term is the DH energy.

\begin{figure}[t]
\includegraphics[width=.99\columnwidth]{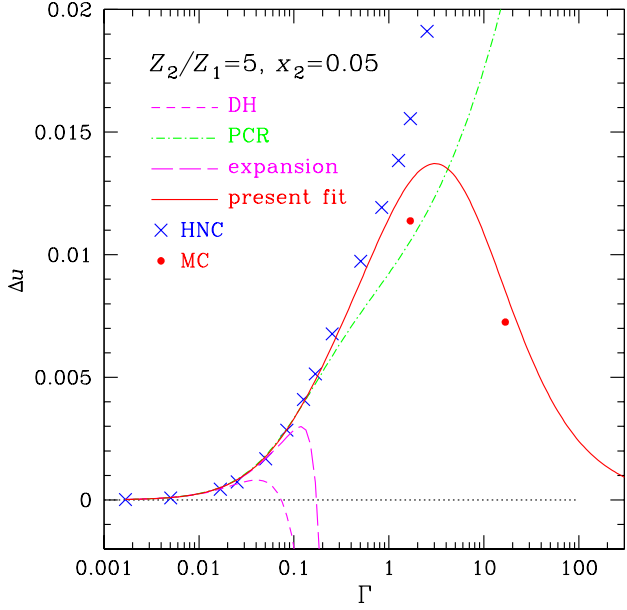}
\caption{\label{fig:umixmc5b}(Color online)
The same as in Fig.~\ref{fig:umixmc2b}
but for $Z_2/Z_1=5$ and $x_2=0.05$.}
\end{figure}

In order to generalize this expression to the case of multicomponent
Coulomb plasmas, let us write the OCP energy in the form
\beq
   u(\Gamma)=\Gamma \tilde{u}(\aion/r_D),
\label{Udecompos}
\eeq
where $\aion$ is the ion sphere radius for the OCP,
and $\tilde{u}$ is the Coulomb energy
per ion in units of $(eZ)^2/a$ ($\tilde{u}=-0.9$
in the ion sphere model \cite{Salpeter54}).
Then the following relation holds
in the DH approximation for multicomponent plasmas:
\begin{equation}
    u=\sum_j x_j \Gamma_j \tilde{u}(\kappa_j),
\label{NLM}
%% \end{equation}
%% %
%% where 
%% \beq
\quad
   \kappa_j \equiv\frac{a_j}{r_D}
   = \sqrt{3\Gamma_e\,\frac{\langle Z^2\rangle}{\langle
   Z\rangle}}
   \,Z_j^{1/3}.
\eeq
Let us assume that relation
(\ref{NLM}) can be applied also
to the higher-order corrections beyond DH.
In this case, according to Eqs.~(\ref{CM}) and~(\ref{Udecompos}),
in the ACM approximation
\bea
   \tilde{u}(\kappa) &=&
            - \frac{\kappa}{2}
           - \kappa^4\,\left[\frac 14\ln\kappa
                          -0.0149085\right]
\nonumber\\&&
           -\kappa^7\,
             \left[\frac{1}{8}\,\ln(\kappa)-0.07369
             \right].
\label{tilde_u}
\eea
Since $\kappa\propto\sqrt{\Gamma}$ for a fixed composition,
$f$ can be obtained from $u$ by integration, which yields
\beq
   f =  \sum_j x_j \Gamma_j \tilde{f}(\kappa_j),
\label{LM_f}
\eeq
where 
\bea
    \tilde{f}(\kappa)  &=& -\frac{\kappa}{3}-
    \frac{\kappa^4}{12}\,(\ln\kappa-0.2263)
\nonumber\\&&
            -\kappa^7 \, (0.02778\,\ln\kappa - 0.01946 ).
\eea

\begin{figure}[t]
\includegraphics[width=.99\columnwidth]{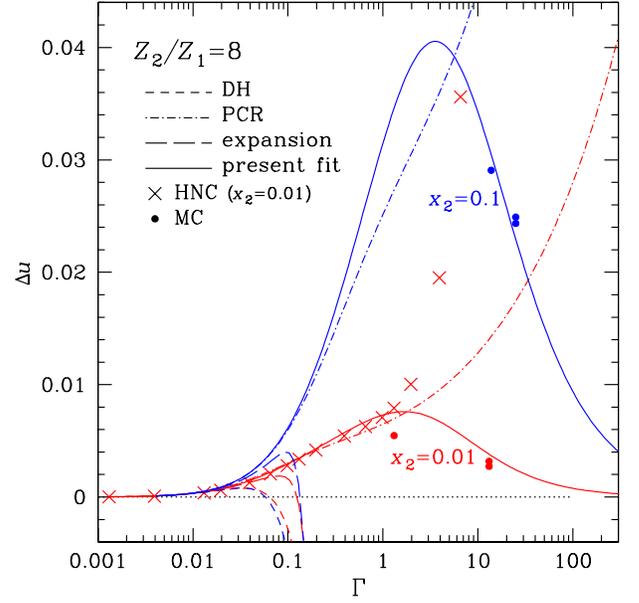}
\caption{\label{fig:umixmc8a}(Color online)
The same as in Fig.~\ref{fig:umixmc2b}
but for $Z_2/Z_1=8$ and two values of $x_2$: 0.01 and 0.1 
(marked near the dots).}
\end{figure}

In Figs.~\ref{fig:umixmc2b}--\ref{fig:umixmc8a} deviations
from the LMR, $\Delta u \equiv u-u_\mathrm{LM}$, calculated according to
Eqs.~(\ref{NLM}) and (\ref{tilde_u}), are plotted by long-dashed lines
and compared to the DH formula (short-dashed lines) and
the HNC data (crosses). 
We see that suggested
approximation (\ref{NLM}) agrees with the data
to much higher $\Gamma$ values than the DH
approximation.

% --------------------------------------------------------
\section{Coulomb liquids at arbitrary coupling}
\label{sect:fit}

In order to find an analytic approximation for the correction
to the LMR in the largest possible interval of
$\Gamma$ for ion gases and liquids, we have selected
from the numerical HNC data \cite{PCR09}
the subset related to $\Gamma\leq1$, which counts
161 different combinations of $x_2$, $Z_2$,
and $\Gamma$
in BIM and 54 combinations of $x_2$, $Z_2$, $x_3$, $Z_3$,
and $\Gamma$
in TIM (assuming $Z_1=1$),
supplemented this HNC data by numerical 
MC data for BIM at $\Gamma>1$ 
(94 combinations of $x_2$, $Z_2$,
and $\Gamma$), and looked for an analytic 
formula which provides a reasonable compromise
between simplicity and accuracy for representing this data.
The MC data has been partly taken from
the previous work \cite{DWSC96,DWS99,DWS03} and partly
obtained by new MC simulations using the same
computer code as before.
Our fitting formula for the
addition to the reduced free energy $f=F/\Nion \kB T$,
relative to the LMR
prediction $f_\mathrm{LM}$,
reads
\beq
   \Delta f\equiv f-f_\mathrm{LM}
 = \frac{\Gamma_e^{3/2} \langle Z^{5/2}\rangle}{\sqrt{3}}
 \,\frac{\delta}{
 (1 + a\, \Gamma^\alpha)\,(1+ b\,\Gamma^\alpha)^\beta},
\label{fit}
\eeq
where $\delta$ is determined by the difference between the LMR
and DH formula at $\Gamma\to0$ (exactly as in Ref.~\cite{PCR09}):
\begin{subequations}
\label{delta}
\beq
   \delta = 1 -
    \frac{\langle Z^2\rangle^{3/2}}{\langle Z\rangle^{1/2}
          \,\langle Z^{5/2}\rangle}
\eeq
for rigid electron background model, and
\beq
   \delta =  \frac{\langle Z\,(Z+1)^{3/2}\rangle}{\langle Z^{5/2}\rangle}
   - \frac{(\langle Z^2\rangle+\langle Z\rangle)^{3/2}
   }{\langle Z\rangle^{1/2}\,\langle Z^{5/2}\rangle}
\label{delta2}
\eeq
\end{subequations}
for polarizable background. 
The expression (\ref{delta2}) for $\delta$ is 
exact in the limit of nondegenerate electrons,
but its use in \req{fit} provides a satisfactory agreement
with numerical data \cite{ChabAsh} obtained 
with allowance for the
polarizability of partially degenerate electron gas
(see \cite{PCR09}).

The fit parameters $a$, $b$, and $\alpha$
are chosen so as to minimize
the mean-square difference between the fit and the data
for
$\Delta u/u_\mathrm{LM}$ at $\Gamma\leq1$
and for
$\Delta u$ at $\Gamma>1$,
while the power index $\beta$ is defined so as 
to quench the increase of $\Delta f$ at $\Gamma\to\infty$.
These parameters
depend on plasma composition as follows:
\bea&&
   a = \frac{2.6\,\delta+14\,\delta^3}{1-\alpha},
\quad
   \alpha = \frac{ \langle Z \rangle^{2/5}}{\langle Z^2 \rangle^{1/5} },
\label{alpha}\\&&
   b = 0.0117\,\left(\frac{\langle Z^2 \rangle }{
    \langle Z \rangle^2}\right)^{\!\!2} a\,,
\quad
   \beta = \frac{3}{2\alpha}-1.
\label{beta}
\eea

The numerical difference of \req{fit}
from the formula in Ref.~\cite{PCR09} 
is small at $\Gamma\lesssim1$, but at
$\Gamma\gg1$ the correction to the LMR prediction
for the reduced internal energy
\beq
   \Delta u = \Gamma\,\frac{\partial(\Delta f)}{\partial\Gamma}
     = \left( \frac32
    - \frac{a\,\alpha\,\Gamma^\alpha}{1 + a \,\Gamma^\alpha}
    - \frac{b\,\alpha\,\beta\,\Gamma^\alpha}{1 + b\, \Gamma^\alpha}
    \right) \Delta f
\label{Deltau}
\eeq
now decreases at large $\Gamma$ in 
agreement with the MC results.
Moreover, \req{Deltau} describes most of the 
data with much higher accuracy than the fit to $\Delta u$ suggested
in Ref.~\cite{DWS03} for BIM at $\Gamma>1$.

A comparison of the numerical
HNC data for $\Delta f$ and $\Delta u$ and MC data for $\Delta u$
to \req{fit} and to the previous fit \cite{PCR09}
shows that the present fit has nearly
the same accuracy as the previous one
for BIM at $\Gamma<1$ (slightly worse for 
small $\Delta u/u$, slightly better for larger $\Delta u/u$),
but it is generally better for TIM at $\Gamma\leq1$ and
substantially better for BIM at $\Gamma>1$.
Examples of $\Gamma$-dependences of $\Delta u$
are shown in Figs.~\ref{fig:umixmc2b}--\ref{fig:umixmc8a}, 
where the dot-dashed lines
correspond to the older fit and the solid lines to the present fit.
The modification of the fit
at small $\Gamma$ values proves to be negligible,
which has been checked
by comparison of fractional differences between the Coulomb
part of the free energy
and the LMR
prediction, as in Ref.~\cite{PCR09},
whereas the modification at large $\Gamma$ can be significant,
as confirmed by Figs.~\ref{fig:umixmc2b}--\ref{fig:umixmc8a}.

\section{Conclusions}
\label{sect:concl}

We have reconsidered free and
internal energies of classical ionic mixtures
in the liquid state, taking into account the results of
HNC calculations in the regime of weak and moderate Coulomb coupling
and MC simulations at strong coupling,
and proposed two new analytic approximations for such mixtures:
the mixing rule (\ref{NLM}), which works well at the
Coulomb coupling parameter $\Gamma<1$, and the
analytic fitting formula
(\ref{fit}), which is along with its derivative (\ref{Deltau})
applicable at any values of $\Gamma$.

\begin{acknowledgments}
The work of A.I.C.\ and A.Y.P.\ was partially supported 
by the Rosnauka Grant NSh-2600.2008.2
and the RFBR Grant 08-02-00837. The work of H.E.D.\ and F.J.R.\ was
partially performed under the auspices of the U.S.\ Department of
Energy by Lawrence Livermore National Laboratory under 
Contract DE-AC52-07NA27344.
\end{acknowledgments}

\end{document}